\documentclass[preprint,showpacs,preprintnumbers,amsmath,amssymb,floatfix]{revtex4}
\usepackage{amsmath}

\usepackage{subfigure}
\usepackage{float}
\usepackage{graphicx}
\usepackage{dcolumn}
\usepackage{bm}
\usepackage{color}
\definecolor{orange}{RGB}{255,127,0}
\definecolor{brown}{RGB}{160,82,45}

\def\undersim#1{\setbox9\hbox{${#1}$}{#1}\kern-\wd9\lower
    2.5pt \hbox{\lower\dp9\hbox to \wd9{\hss $_\sim$\hss}}}
\def\undersim#1{\setbox9\hbox{${#1}$}{#1}\kern-\wd9\lower
    2.5pt \hbox{\lower\dp9\hbox to \wd9{\hss $_\sim$\hss}}}

\def\mr{{\mathbf r}}
\def\mK{{\mathbf K}}
\def\mr{{\mathbf r}}
\def\mq{{\mathbf q}}
\def\mk{{\mathbf p}}

\begin{document}

\title{Two-pion interferometry for partially coherent sources in relativistic
heavy-ion collisions in a multiphase transport model}

\author{Shi-Yao Wang, Jun-Ting Ye, Wei-Ning Zhang\footnote{wnzhang@dlut.edu.cn}}
\affiliation{School of Physics, Dalian University of Technology, Dalian, Liaoning
116024, China}


\begin{abstract}
We perform two-pion Hanbury Brown-Twiss (HBT) interferometry for the partially
coherent pion-emitting sources in relativistic heavy-ion collisions, using a
multi-phase transport (AMPT) model. A longitudinal coherent emission length, as
well as a transverse coherent emission length, are introduced to the pion generation
coordinates in calculating the HBT correlation functions of the partially coherent
sources. We compare the model results with and without coherent emission
conditions with experimental data in Au-Au collisions at center-of-mass energy
$\sqrt{s_{NN}}=$200 GeV, and in Pb-Pb collisions at center-of-mass energy
$\sqrt{s_{NN}}=$2.76 TeV, and find that the HBT results of the partially coherent
sources are closer to the experimental data than those of chaotic sources.
\end{abstract}


\maketitle

\section{Introduction}
Two-pion Hanbury Brown-Twiss (HBT) correlation is an important observable to detect
the space-time structure of particle-emitting sources produced in relativistic
heavy-ion collisions \cite{Gyu79,Wongbook94,Wie99,Wei00,Csorgo02,Lisa05}.
Because the two pions emitted coherently have no HBT correlation, this observable
can also be used to study the source coherence
\cite{Gyu79,Wongbook94,Wie99,Wei00,Csorgo02,Lisa05}.
Recent measurements of two-pion HBT correlations in Pb-Pb collisions at the Large
Hadron Collider (LHC) \cite{ALICE_PRC89_2014,ALICE_PRC92_2015,ALICE_PRC93_2016a} and
measurements in Au-Au collisions at the Relativistic Heavy Ion Collider (RHIC)
\cite{STAR_PRL87_2001,PHENIX_PRL93_2004,STAR_PRC71_2005,STAR_PRC81_2010} showed that
the two-pion correlation functions near zero relative momentum are substantially
less than 2, corresponding to the magnitude for completely chaotic sources
\cite{Gyu79,Wongbook94,Wie99,Wei00,Csorgo02,Lisa05}. These observations and the
suppression of the multi-pion correlation functions observed in Pb-Pb collisions
at the LHC \cite{ALICE_PRC89_2014,ALICE_PRC93_2016} indicate that the pion-emitting
sources are perhaps partially coherent. However, there is no widely accepted explanation
for these observations.

Because of the complexity of relativistic heavy-ion collision systems, the calculations
for experimental observables based on models play an important role in understanding
the system properties. A multi-phase transport model (AMPT) has been extensively used
in relativistic heavy-ion collisions \cite{{LinKo_PRC65_2002,LinKo_JPG30_2004,
Lin_PRC72_2005,Nasim_PRC82_2010,XJunCMKo_PRC83_2011,DSolanki_PLB720_2013,
YLXie_NPA920_2013,BzdakGLMa_PRL113_2014,GLMaZWLin_PRC93_2016,HLi_PRC96_2017,
Haque_JPG46_2019,MDordevic_PRC101_2020,TShao_PRC102_2020,KJSunCMKo_PRC103_2021,
MagdtLacey_PRC104_2021,SBasu_PRC104_2021,MagEvdLac_JPG48_2021,ZZhangNYuHXu_EPJA58_2022,
BaryZhangRuYang_CPC_2021}}. In Ref. \cite{SYWangWNZhang_FP_2022}, the authors studied 
the two-pion HBT correlation functions for the partially coherent pion-emitting sources 
constructed with the AMPT model by introducing a coherent emission length $L_{\rm C}$ 
to the pion longitudinal freeze-out coordinates. It was assumed that the emission of 
the two pions with a longitudinal difference of freeze-out coordinates smaller than 
$L_{\rm C}$ is coherent, and otherwise it is completely chaotic \cite{SYWangWNZhang_FP_2022}. They found that the intercepts of the two-pion correlation 
functions of the partially coherent sources with finite coherent emission lengths are 
less than those of the completely chaotic sources in Au-Au collisions at center-of-mass 
energy $\sqrt{s_{NN}}=$200~GeV, and in Pb-Pb collisions at center-of-mass energy 
$\sqrt{s_{NN}}=$2.76~TeV.

As a microscopic transport model, the AMPT model can provide the coordinates and momenta
of the freeze-out particles at each time step in the source after they are generated.
In this paper, we introduce a transverse coherent emission length and a longitudinal
coherent emission length, which are momentum dependent according to particle de Broglie
wavelength, to identical pion generation coordinates to construct the partially coherent
pion-emitting sources. We calculate the two-pion correlation functions for the
partially coherent sources in the AMPT model, and compare the model results with
experimental data in Au-Au collisions at $\sqrt{s_{NN}}=$200~GeV \cite{STAR_PRC71_2005}
and in Pb-Pb collisions at $\sqrt{s_{NN}}=$2.76~TeV \cite{ALICE_PRC93_2016a}.
We find that the HBT results of partially coherent sources are closer to experimental
data than those of chaotic sources. The results of coherent fraction for the partially
coherent sources in Pb-Pb collisions are consistent with the values extracted from the
experimental measurements of multi-pion HBT correlations
\cite{ALICE_PRC89_2014,ALICE_PRC93_2016}.

The rest of this paper is organized as follows. Section II presents a brief
introduction to the AMPT model, and discusses the generation and freeze-out coordinates
of identical pions generated by quark coalescence and by particle scattering and
decay. Section III gives the two-pion correlation function
calculations for the partially coherent and completely chaotic sources. Model HBT
interferometry results in Au-Au collisions at $\sqrt{s_{NN}}=$200~GeV, and in Pb-Pb
collisions at $\sqrt{s_{NN}}=$2.76~TeV, are presented and compared with experimental
data at the RHIC and LHC. We summarize and discuss this work in section IV.

\section{Pion generation and freeze-out in AMPT model}
The AMPT model is a hybrid composed of initialization, parton transport, hadronization,
and hadron transport \cite{Lin_PRC72_2005}. It has been successfully used
to describe the observables in high-energy heavy-ion collisions at the RHIC and LHC
\cite{{LinKo_PRC65_2002,LinKo_JPG30_2004,Lin_PRC72_2005,Nasim_PRC82_2010,
XJunCMKo_PRC83_2011,DSolanki_PLB720_2013,YLXie_NPA920_2013,BzdakGLMa_PRL113_2014,
GLMaZWLin_PRC93_2016,HLi_PRC96_2017,Haque_JPG46_2019,MDordevic_PRC101_2020,
TShao_PRC102_2020,KJSunCMKo_PRC103_2021,MagdtLacey_PRC104_2021,SBasu_PRC104_2021,
MagEvdLac_JPG48_2021,ZZhangNYuHXu_EPJA58_2022}}.
The initialization of collisions in the AMPT model is performed using the HIJING
model \cite{HIJING}. The parton and hadron transport are described by
the ZPC (Zhang's parton cascade) model \cite{ZPC} and ART model \cite{ART}, respectively.
In this study, we use the string melting version of the AMPT model in which partons
are hadronized by the quark coalescence mechanism \cite{LinKo_PRC65_2002,Lin_PRC72_2005}.
We take the model parameter $\mu$ of parton screening mass to be 2.2814~fm$^{-1}$, and
the strong coupling constant $\alpha_s$ to be 0.47, corresponding to a
parton-scattering cross section of 6~mb \cite{Lin_PRC72_2005}.

\vspace*{0mm}
\begin{figure}[htbp]
\includegraphics[scale=0.76]{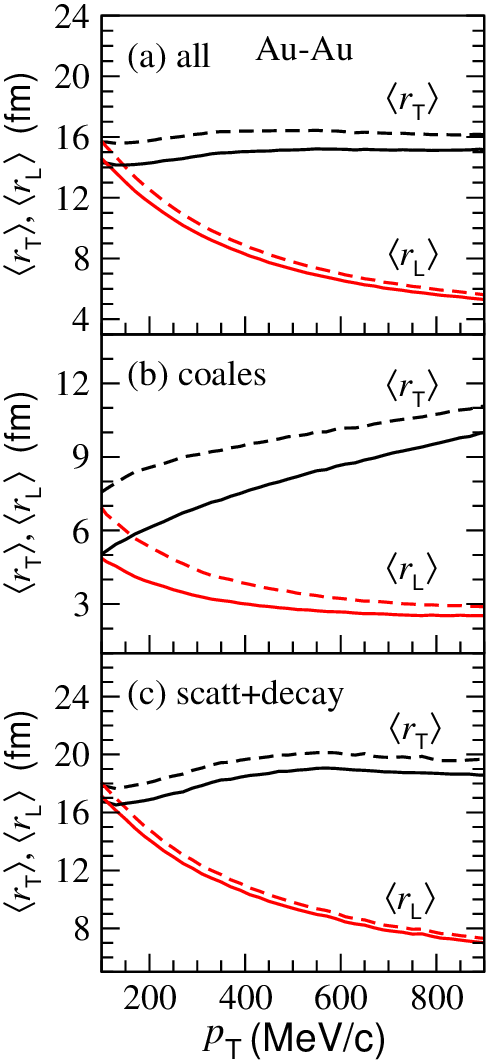}
\caption{(Color online) Pion average transverse and longitudinal freeze-out coordinates
(dashed lines) and generation coordinates (solid lines) in Au-Au collisions at
$\sqrt{s_{NN}}=$200~GeV. (a) all identical (negative) pions; (b) the pions generated by
quark coalescence; (c) the pions generated by particle scattering and decay.}
\label{zf-rtrl-pt-RHIC}
\end{figure}

\vspace*{0mm}
\begin{figure}[htbp]
\includegraphics[scale=0.76]{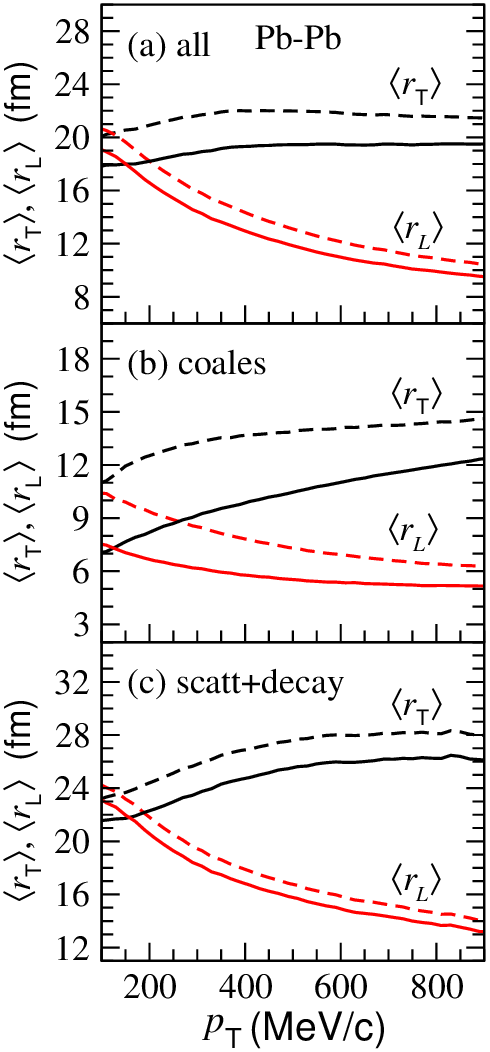}
\caption{(Color online) Pion average transverse and longitudinal freeze-out coordinates
(dashed lines) and generation coordinates (solid lines) in Pb-Pb collisions at
$\sqrt{s_{NN}}=$2.76~TeV. (a) all the identical (negative) pions; (b) the pions generated
by quark coalescence; (c) the pions generated by particle scattering and decay.}
\label{zf-rtrl-pt-LHC}
\end{figure}

After hadronization, the generated hadrons may interact with other particles in the
source, and may be absorbed or remain until freeze-out. Some
hadrons may also be generated by particle interactions in the source and hadron
decays. With the AMPT model, we can trace back to the origin of a freeze-out particle,
its generation coordinates, momentum, and parent.
For instance, we can track back step-by-step for a recorded pion with freeze-out
coordinate $\mr_f$ to find its first appeared position $\mr_g$, called generation coordinates, according to the recording files of the AMPT model.
Figure~\ref{zf-rtrl-pt-RHIC} shows the average
transverse and longitudinal freeze-out coordinates (dashed lines) and generation
coordinates (solid lines) of identical pions versus the particle transverse momentum
in Au-Au collisions at $\sqrt{s_{NN}}=$200~GeV, where panels (a) -- (c) are for all
identical (negative) pions, the pions generated by quark coalescence, and the pions
generated by particle scattering and hadron decays, respectively.
In our calculations, the cutoff time for hadron cascade is taken to be 200~fm$/c$,
as in previous HBT studies \cite{Lin_PRC72_2005}. The impact parameter is taken
to be 0--3~fm. For Au-Au collisions at $\sqrt{s_{NN}}=$200~GeV, we take a rapidity
cut, $|y|<0.5$, as in experimental data analyses \cite{STAR_PRC71_2005}.
It can be seen that the average freeze-out coordinates are larger than the average
generation coordinates. The average values for quark coalescence generation are the
smallest, and the average values for the pions generated by particle scattering and
decay are the largest. This is because pions generated by quark coalescence occur
earlier on average than those generated by particle scattering and decay. One can
also see that the relative difference between the average freeze-out coordinate and
generation coordinate for quark coalescence is larger than that for particle scattering
and decay.

Figure~\ref{zf-rtrl-pt-LHC} shows the average transverse and longitudinal freeze-out
coordinates (dashed lines) and generation coordinates (solid lines) of identical pions
versus the particle transverse momentum in Pb-Pb collisions at $\sqrt{s_{NN}}=$2.76~TeV,
where panel (a) is for all of the identical (negative) pions, (b) is for the pions
generated by quark coalescence, and (c) is for the pions generated by particle scattering
and hadron decays. In the calculations for Pb-Pb collisions, we take a pseudorapidity cut,
$|\eta|<0.8$, as in experimental data analyses \cite{ALICE_PRC93_2016a}. One can see that
the average coordinates are larger in Pb-Pb collisions at $\sqrt{s_{NN}}=$2.76~TeV
than in Au-Au collisions at $\sqrt{s_{NN}}=$200~GeV.

\section{Two-pion HBT analyses}
\subsection{Calculation of two-pion correlation function in AMPT model}
The two-pion HBT correlation function is defined as:
\begin{equation}
C(\mk_1,\mk_2)=\frac{P(\mk_1,\mk_2)}{P(\mk_1)P(\mk_2)},
\label{eq-Cp}
\end{equation}
where $P(\mk_i)~(i=1,2)$ is the distribution of single-pion momentum $\mk_i$, and
$P(\mk_1,\mk_2)$ is the two identical pion momentum distribution in an event.

For a chaotic pion-emitting source, the denominator and numerator in Eq.~(\ref{eq-Cp})
can be expressed respectively as \cite{Wongbook94}
\begin{equation}
\label{Ppi}
P(\mk_1)P(\mk_2)=\sum_{X_1,X_2} A^2(\mk_1,X_1) A^2(\mk_2,X_2),
\end{equation}
\begin{equation}
\label{Pp1p2}
P(\mk_1,\mk_2)=\sum_{X_1,X_2} \left|\Phi(\mk_1,\mk_2;X_1,X_2)\right|^2,
\end{equation}
where $A(\mk_i,X_i)$ is the magnitude of the amplitude for emitting a pion with
momentum $\mk_i$ at four-coordinate $X_i$ (freeze-out coordinates), and
\begin{equation}
\Phi(\mk_1,\mk_2;X_1,X_2)=\frac{1}{\sqrt{2}}\Big[A(\mk_1,X_1)A(\mk_2,X_2)e^{ip_1\cdot
X_1} e^{ip_2\cdot X_2} +A(\mk_1,X_2)A(\mk_2,X_1)e^{ip_1\cdot X_2} e^{ip_2\cdot X_1} \Big],
\end{equation}
where $p_i$ is the four-momentum.

Because the HBT effect works on the identical pion pair with small relative momentum
$\mq=\mk_1\!-\!\mk_2$, we can substitute $A(\mk_j,X_i)$ for $A(\mK,X_i)$ $(i,j=1,2)$
[smoothed approximation, $\mK=(\mk_1+\mk_2)/2$]. Then we have
\begin{eqnarray}
P(\mk_1,\mk_2)&=&\sum_{X_1,X_2}A^2(\mK,X_1)A^2(\mK,X_2)\left\{1+\cos\left[(p_1-p_2)\cdot
(X_1-X_2)\right]\right\}
\nonumber\\
&=&\sum_{X_1,X_2}A^2(\mk_1,X_1)A^2(\mk_2,X_2)\left\{1+\cos\left[(p_1-p_2)\cdot
(X_1-X_2)\right]\right\}.
\end{eqnarray}

In relativistic heavy-ion collisions at the RHIC and LHC, the systems are initially
compressed in the beam direction (longitudinal or $z$ direction), and then expand
longitudinally and transversely. A partially coherent pion-emitting source was constructed
in the AMPT model by introducing a coherent emission length $L_{\rm C}$ to the longitudinal
freeze-out coordinates of identical pions \cite{SYWangWNZhang_FP_2022}. It is assumed that
the emissions of the two pions are coherent if they have a longitudinal distance of freeze-out coordinates less than $L_{\rm C}$. However, a constant coherent emission
length is too simple.
Pions, as the lightest hadron, have wide de Broglie wavelengths, $(h/|\mk|)$. Considering
the systems produced in relativistic heavy-ion collisions are anisotropic in longitudinal
and transverse directions, we respectively introduce longitudinal and transverse coherent
emission lengths as:
\begin{equation}
L_{\rm CZ}=a_Z\left[\frac{h}{k_{1Z}}+\frac{h}{k_{2Z}}\right],
\label{eq-LZ}
\end{equation}
and
\begin{equation}
L_{\rm CT}=a_T\left[\frac{h}{k_{1T}}+\frac{h}{k_{2T}}\right],
\label{eq-LT}
\end{equation}
where $a_Z$ and $a_T$ are two coherent-length parameters determined by comparing model HBT
results with experimental data, $k_{iZ}$ and $k_{iT}$ ($i=1,2$) are pion longitudinal and
transverse generation momenta. We assume that the emission of two pions with differences
of longitudinal and transverse generation coordinates respectively smaller than $L_{\rm CZ}$ and $L_{\rm CT}$ are coherent.
For this partially coherent source, we have
\begin{equation}
P(\mk_1,\mk_2)=\sum_{X_1,X_2}A^2(\mk_1,X_1)A^2(\mk_2,X_2)R(\mk_1,\mk_2;X_1,X_2),
\end{equation}
where
\begin{equation}
R(\mk_1,\mk_2;X_1,X_2)=1, \hspace*{36mm}{\rm for}~\Delta z<L_{\rm CZ}~{\rm and}~
\Delta r_{\rm T}<L_{\rm CT},
\end{equation}
and
\begin{equation}
R(\mk_1,\mk_2;X_1,X_2)=1+\cos\left[(p_1-p_2)\cdot (X_1-X_2)\right], ~~~~~~~~
{\rm otherwise}, \hspace*{12mm}
\end{equation}
where $\Delta z=|z_1-z_2|$, $\Delta r_{\rm T}=\sqrt{(x_1-x_2)^2+(y_1-y_2)^2}$, and
$(x_i,y_i,z_i)~(i=1,2)$ are the pion generation coordinates.

\begin{figure}[htbp]
\includegraphics[scale=0.76]{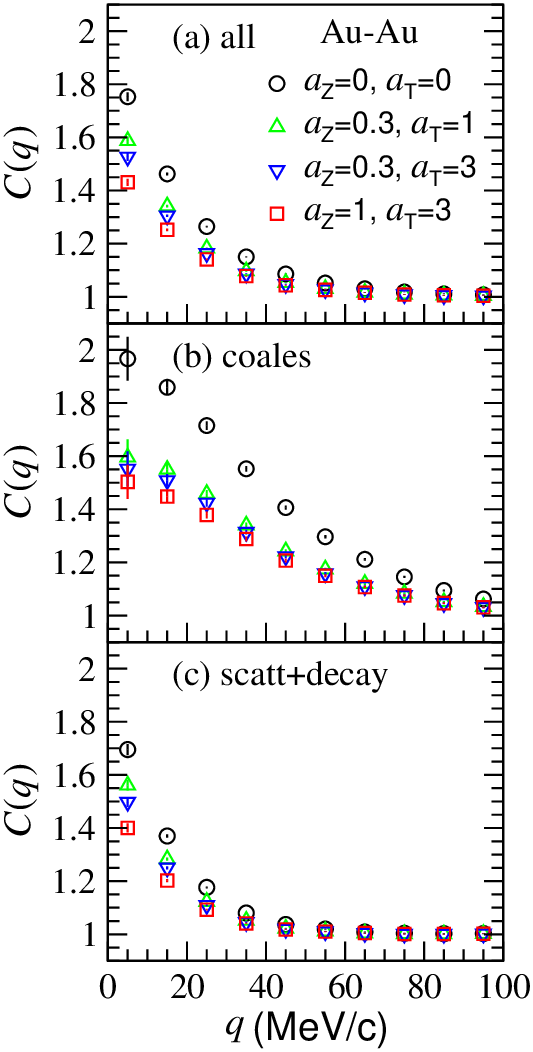}
\caption{(Color online) Two-pion correlation functions $C(q)$ for pion-emitting sources
with different coherent-length parameters, ($a_{\rm Z}=0$, $a_{\rm T}=0$), ($a_{\rm Z}=
0.3$, $a_{\rm T}=1$), ($a_{\rm Z}=0.3$, $a_{\rm T}=3$), and ($a_{\rm Z}=1$, $a_{\rm T}=
3$), in Au-Au collisions at $\sqrt{s_{NN}}=200$~GeV, with impact parameter $0<b<3$~fm
in the AMPT model. Results calculated with (a) all of the identical (negative) pions;
(b) the pions generated by quark coalescence; (c) the pions generated by particle
scattering and decay.}
\label{zf_CqAu}
\end{figure}

\begin{figure}
\includegraphics[scale=0.76]{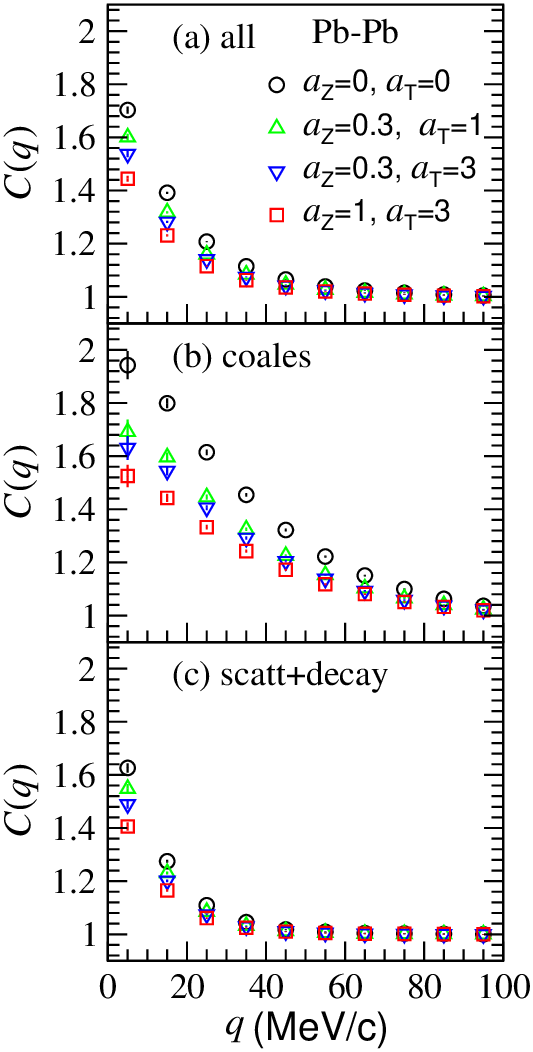}
\caption{(Color online) Two-pion correlation functions $C(q)$ for pion-emitting sources
with different coherent-length parameters, ($a_{\rm Z}=0$, $a_{\rm T}=0$), ($a_{\rm Z}=
0.3$, $a_{\rm T}=1$), ($a_{\rm Z}=0.3$, $a_{\rm T}=3$), and ($a_{\rm Z}=1$, $a_{\rm T}=
3$), in Pb-Pb collisions at $\sqrt{s_{NN}}=2.76$~TeV, with impact parameter $0<b<3$~fm
in the AMPT model. Results calculated with (a) all of the identical (negative) pions;
(b) the pions generated by quark coalescence; (c) the pions generated by particle
scattering and decay.}
\label{zf_CqPb}
\end{figure}

Figures~\ref{zf_CqAu}(a)--(c) show the two-pion
correlation functions $C(q)$ calculated with all of the identical (negative) pions, the
pions generated by quark coalescence, and by particle scattering and decay, respectively,
in Au-Au collisions at $\sqrt{s_{NN}}=$200~GeV and with impact parameter $0<b<3$~fm
in the AMPT model. The correlation functions for the pions generated by quark coalescence
are wider than those for the pions generated by particle scattering and decay, because
the former have smaller average freeze-out coordinates (see Fig.~\ref{zf-rtrl-pt-RHIC}).
In Figs.~\ref{zf_CqAu}(a)--(c), the circle, up-triangle, down-triangle, and square symbols denote the coherent-length parameters ($a_{\rm Z}=0$, $a_{\rm T}=0$), ($a_{\rm Z}=0.3$,
$a_{\rm T}=1$), ($a_{\rm Z}=0.3$, $a_{\rm T}=3$), and ($a_{\rm Z}=1$, $a_{\rm T}=3$),
respectively. One can see that the correlation function values for the partially coherent
sources are less than those for completely chaotic sources ($a_{\rm Z}=a_{\rm T}=0$) at
small relative momenta. Also, one can see that the difference between the correlation
functions of partially coherent source and chaotic source at small $q$ is greater for
the quark coalescence generation than that for the generation of particle scattering
and decay. This is because that the smaller average generation coordinates for quark
coalescence generation lead to a greater effect than that for the generation of particle
scattering and decay, with the same coherent-length parameters.

Figures~\ref{zf_CqPb}(a)--(c) respectively show the two-pion
correlation functions $C(q)$ calculated with all of the identical (negative) pions, the
pions generated by quark coalescence, and the pions generated by particle scattering and decay, in Pb-Pb collisions at $\sqrt{s_{NN}}=$2.76~TeV and with impact parameter
$0<b<3$~fm in the AMPT model, where the circle, up-triangle, down-triangle, and square 
symbols mean the same as in Fig. \ref{zf_CqAu}. Compared to the correlation functions 
in Au-Au collisions, the two-pion correlation functions in Pb-Pb collisions are narrower 
because the average source sizes in Pb-Pb collisions are larger.

\subsection{Interferometry results}
In two-pion HBT analyses, the fitting correlation function is usually the Gaussian form,
\begin{equation}
C(q_{\rm out},q_{\rm side},q_{\rm long})=\kappa (1+\lambda \,e^{-q^2_{\rm out}R^2_{\rm
out} -q^2_{\rm side}R^2_{\rm side} -q^2_{\rm long}R^2_{\rm long}}),
\label{Cqosl}
\end{equation}
where $q_{\rm out}$, $q_{\rm side}$, and $q_{\rm long}$ are the Bertsch-Pratt
variables \cite{BerGonToh_PRC37_NPA498,PraCsoZim_PRC42}, which respectively denote the
components of the relative momentum $\mq$ in the transverse ``out" (parallel to the
transverse momentum of the pion pair $\mK_{\rm T}$), transverse ``side" (in the transverse
plane and perpendicular to $\mK_{\rm T}$), and longitudinal (``long") directions in the
longitudinally co-moving system (LCMS) frame \cite{Lisa05}; $\kappa$ is a normalization
parameter, $\lambda$ is the chaoticity parameter for the partially coherent source; and $R_{\rm out}$, $R_{\rm side}$, and $R_{\rm long}$ are the HBT radii in the respective out,
side, and long directions.

\vspace*{5mm}
\begin{figure}[htbp]
\includegraphics[scale=0.7]{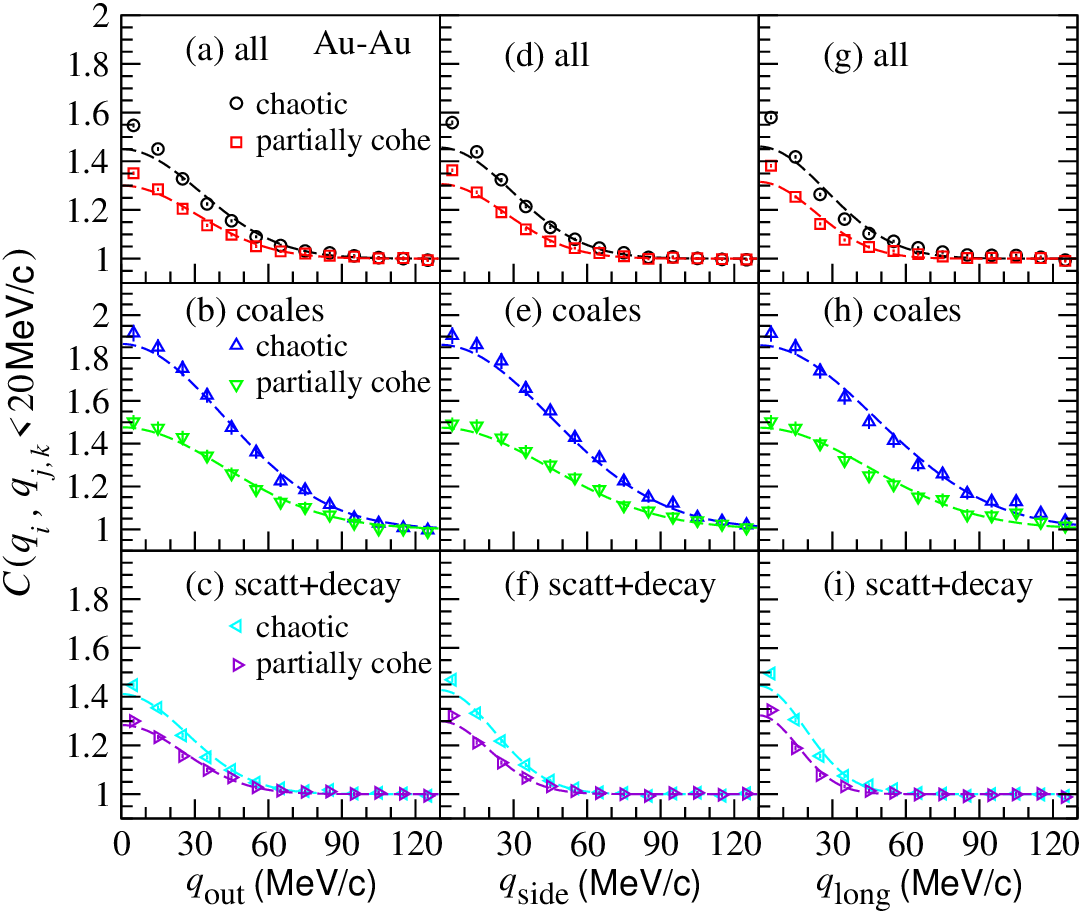}
\caption{(Color online) Two-pion correlation functions with respect to relative
momenta $q_{i,j,k}$ ($i,j,k={\rm out, side, long}$) for chaotic and partially
coherent sources in Au-Au collisions at $\sqrt{s_{NN}}=200$~GeV, with impact
parameter $0<b<3$~fm in the AMPT model. Top: all of the pions; middle: pions
generated by quark coalescence; bottom: pions generated by particle scattering
and decay.}
\label{zf-Cqosl-au}
\end{figure}

Figure~\ref{zf-Cqosl-au} shows projections of the two-pion correlation functions
$C(q_{\rm out},q_{\rm side},q_{\rm long})$ in the out, side, and long directions for
chaotic and partially coherent sources in Au-Au collisions at $\sqrt{s_{NN}}=
200$~GeV, with impact parameter $0<b<3$~fm in the AMPT model, where, the top panels
are for all of the identical pions, middle panels are for the pions generated by
quark coalescence, and bottom panels are for the pions generated by particle
scattering and decay. For each relative momentum direction, the projections of the
correlation functions are obtained from the three-dimensional correlation functions
by averaging the relative momenta in the other two directions over 0 to 20~MeV/$c$.
Dashed lines in the figure are the fitting curves of Eq.~(\ref{Cqosl}).
In the calculations, the longitudinal and transverse coherent-length parameters for
the partially coherent sources are taken as $a_{\rm Z}=0.5$ and $a_{\rm T}=1.8$,
respectively, according to the comparison of the fitted HBT results in the AMPT model
with experimental data \cite{STAR_PRC71_2005} in different intervals of transverse
momentum of the pion pair. The total number of events in the model calculations is 
$3\times10^4$.

\vspace*{5mm}
\begin{figure}[htbp]
\includegraphics[scale=0.7]{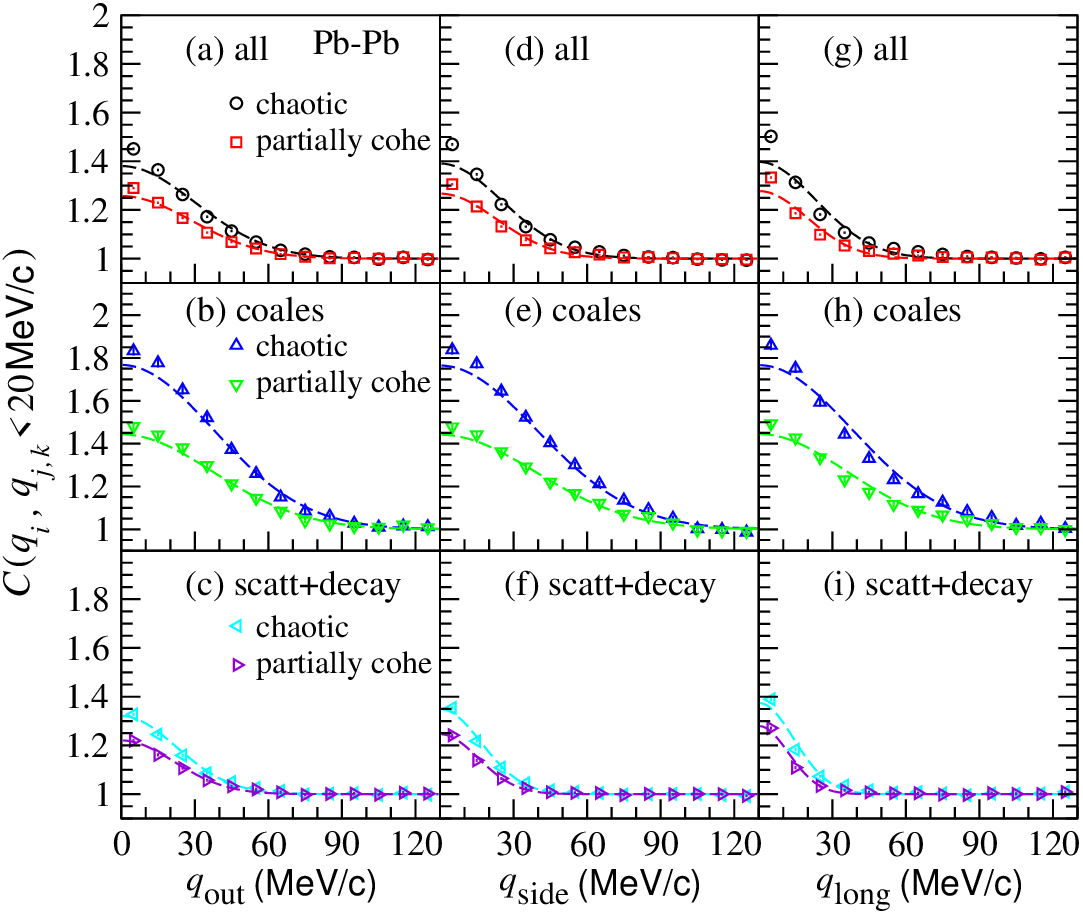}
\caption{(Color online) Two-pion correlation functions with respect to relative
momenta $q_{i,j,k}$ ($i,j,k={\rm out, side, long}$) for chaotic and partially
coherent sources in Pb-Pb collisions at $\sqrt{s_{NN}}=2.76$~TeV, with impact
parameter $0<b<3$~fm in the AMPT model. Top: all of the pions; middle: pions
generated by quark coalescence; bottom: pions generated by particle scattering
and decay.}
\label{zf-Cqosl-pb}
\end{figure}

Figure~\ref{zf-Cqosl-pb} shows projections of the two-pion correlation functions
$C(q_{\rm out},q_{\rm side},q_{\rm long})$ in the out, side, and long directions for
chaotic and partially coherent sources in Pb-Pb collisions at $\sqrt{s_{NN}}=
2.76$~TeV, with impact parameter $0<b<3$~fm in the AMPT model. In calculations in
Pb-Pb collisions, the longitudinal and transverse coherent-length parameters for the
partially coherent sources are taken to be $a_{\rm Z}=0.8$~fm and $a_{\rm T}=2.5$,
respectively, according to the comparison of the fitted HBT results in the AMPT model
with experimental data \cite{ALICE_PRC93_2016a} in different intervals of transverse
momentum of the pion pair. The total number of events in the mode calculations in 
Pb-Pb collisions is $6\times10^3$.

One can see that the intercepts of the correlation functions of partially coherent
sources at zero relative momentum are lower than those of chaotic sources. This
effect is most significant for the pions generated by quark coalescence and the least
significant for the pions generated by particle scattering and decay. This is because
the average generation coordinate of pions generated by quark coalescence is the
smallest, and that of pions generated by particle scattering and decay is the largest.

\begin{figure}[htbp]
\includegraphics[scale=0.7]{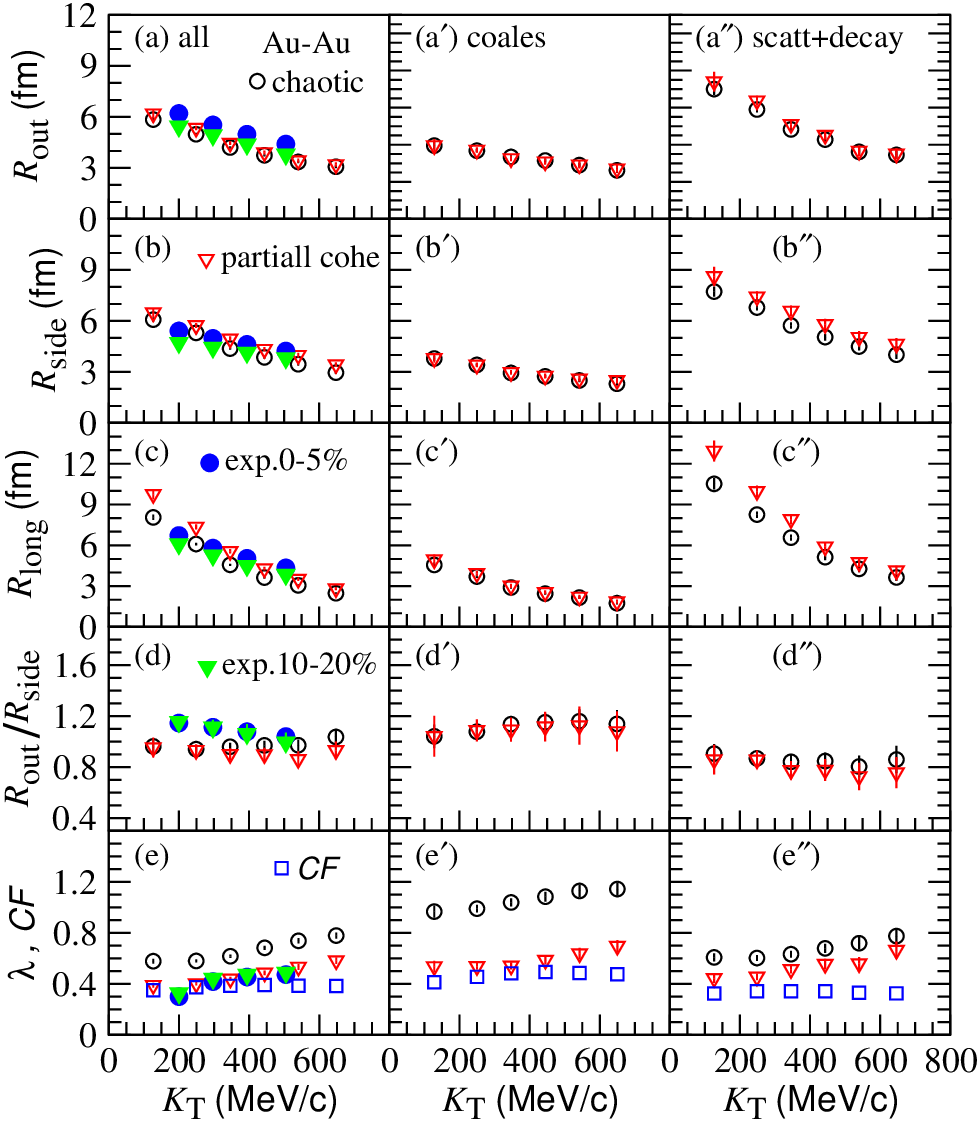}
\caption{(Color online) Two-pion HBT results ($R_{\rm out}$, $R_{\rm side}$, $R_{\rm
long}$, $R_{\rm out}/R_{\rm side}$, $\lambda$) for the chaotic and partially coherent
source and the coherent fraction ($C\!F$) of partially coherent source, with respect
to transverse momentum of pion pair $K_{\rm T}$, in Au-Au collisions at
$\sqrt{s_{NN}}=200$~GeV with impact parameter $0<b<3$~fm in the AMPT model.
Experimental data \cite{STAR_PRC71_2005} in Au-Au collisions at
$\sqrt{s_{NN}}=200$~GeV with centralities 0-5\% and 10-20\% are presented for
comparison.}
\label{zf-HBTfit-au}
\end{figure}

Figure~\ref{zf-HBTfit-au} shows the results of the fitted HBT radii $R_{\rm out}$,
$R_{\rm side}$, and $R_{\rm long}$; ratio $R_{\rm out}/R_{\rm side}$; and
chaoticity parameter $\lambda$ for chaotic and partially coherent sources in
AMPT Au-Au collisions at $\sqrt{s_{NN}}=200$~GeV, with impact parameter $0<b<3$~fm,
in the transverse momentum intervals of pion pair $K_{\rm T}<200$~MeV$\!/c$, $200\leq
K_{\rm T}<300$~MeV$\!/c$, $300\leq K_{\rm T}<400$~MeV$\!/c$, $400\leq K_{\rm T}<500
$~MeV$\!/c$, $500\leq K_{\rm T}<600$~MeV$\!/c$, and $600\leq K_{\rm T}<1000$~MeV$ \!/c$.
The experimental data in Au-Au collisions at $\sqrt{s_{NN}}=200$~GeV with centralities
0-5\% and 10-20\% \cite{STAR_PRC71_2005} are also shown in
Figs.~\ref{zf-HBTfit-au}(a)--(e) for comparison.
We found that increasing the longitudinal coherent-length parameter $a_{\rm Z}$ leads
to a decrease of $\lambda$ and increase of $R_{\rm long}$, while increasing the
transverse coherent-length parameter $a_{\rm T}$ decreases $\lambda$ and increases
$R_{\rm side}$ and $R_{\rm out}$. However, the increase of $a_{\rm T}$ does not improve
the results of $R_{\rm out}/R_{\rm side}$. Finally, we took the longitudinal and
transverse coherent-length parameters $a_{\rm Z}$ and $a_{\rm T}$ to be 0.5 and 1.8,
respectively, for the partially coherent source. One can see that the model HBT radii
and chaoticity parameter of the partially coherent source are closer to the experimental
data compared to those of the chaotic source.
The middle and right panels in Fig.~\ref{zf-HBTfit-au} show the results for chaotic
and partially coherent sources with the pions generated by quark coalescence and
particle scattering and decay, respectively. One can see that the results of $\lambda$
for quark coalescence show great differences between chaotic and partially coherent
sources. However, the corresponding differences of $\lambda$ values are small in the
case of particle scattering and decay.

\begin{table*}[htb]
\begin{center}
\caption{Ratio of coherent pion-pair number $n_c/n_t$, and coherent fraction $C\!F$, in
Au-Au collisions at at $\sqrt{s_{NN}}=200$~GeV, in the AMPT model. The superscripts (1),
(2), and (3) are for all of the generated pions, the pions generated by quark coalescence,
and the pions generated by particle scattering and decay, respectively.}
\begin{tabular}{c|cc|cc|cc}
\hline\hline
Au-Au@200\,GeV&~$n_c^{(1)}/n_t^{(1)}$~&~~$C\!F^{(1)}$~~&~~$n_c^{(2)}/n_t^{(2)}$~&
~~$C\!F^{(2)}$~~&~~$n_c^{(3)}/n_t^{(3)}$~&~~$C\!F^{(3)}$~~\\
\hline
$K_T\!<\!200\,\text{MeV\!/\!c}$&0.122&0.350&0.170&0.413&0.106&0.325\\
~$200\!\!\leq\!K_T\!<\!\!300\,\text{MeV\!/\!c}$&0.140&0.374&0.207&0.455&0.117&0.343\\
~$300\!\!\leq\!K_T\!<\!\!400\,\text{MeV\!/\!c}$&0.148&0.385&0.233&0.483&0.118&0.343\\
~$400\!\!\leq\!K_T\!<\!\!500\,\text{MeV\!/\!c}$&0.153&0.391&0.244&0.494&0.117&0.342\\
~$500\!\!\leq\!K_T\!<\!\!600\,\text{MeV\!/\!c}$&0.149&0.386&0.236&0.486&0.110&0.331\\
~$600\!\!\leq\!K_T\!<\!\!1000\,\text{MeV\!/\!c}$&0.147&0.384&0.226&0.476&0.106&0.326\\
\hline\hline
\end{tabular}
\label{Tab-coh-Au}
\end{center}
\end{table*}

With the numbers of coherent-emission pion-pairs $n_c=N_c(N_c-1)/2$ and total pion-pairs $n_t=N_t(N_t-1)/2$, where $N_c$ and $N_t$ are the corresponding pion numbers,
we can obtain the source coherent fraction $C\!F=N_c/N_t$.
In Figs.~\ref{zf-HBTfit-au}(e), \ref{zf-HBTfit-au}(e$'$), and \ref{zf-HBTfit-au}(e$''$),
the square symbols give the results of coherent fraction for the partially coherent
source in Au-Au collisions at $\sqrt{s_{NN}}=200$~GeV in the AMPT model. At small
transverse momenta of pion pair, the values of coherent fraction are mainly determined
by pion longitudinal momentum. However, the decrease of coherent fraction with increasing
$K_{\rm T}$ at large transverse momenta of pion pair is mainly determined by the increase
of pion transverse momentum at large transverse momenta of pion pair, which leads to a
decrease of $L_{\rm CT}$ [see Eq.~(\ref{eq-LT})].
Table I presents the ratio of coherent pion-pair number to total pion-pair number,
$n_c/n_t$, and the corresponding coherent fraction, $C\!F$, where the superscripts (1),
(2), and (3) are for the ``all", ``coales", and ``scatt+decay" cases in
Fig.~\ref{zf-HBTfit-au}, respectively. One can see that the results of
$n_c^{(2)}/n_t^{(2)}$ are largest and the results of $n_c^{(3)}/n_t^{(3)}$ are smallest,
because the average generation coordinate of the pions generated by quark coalescence
is smallest and the average generation coordinate of the pions generated by particle
scattering and decay is largest. This leads to the largest results of $C\!F^{(2)}$ and
smallest results of $C\!F^{(3)}$. The result in the ``(1)" case is between the results
in the ``(2)" and ``(3)" cases.

\vspace*{0mm}
\begin{figure}[htbp]
\includegraphics[scale=0.7]{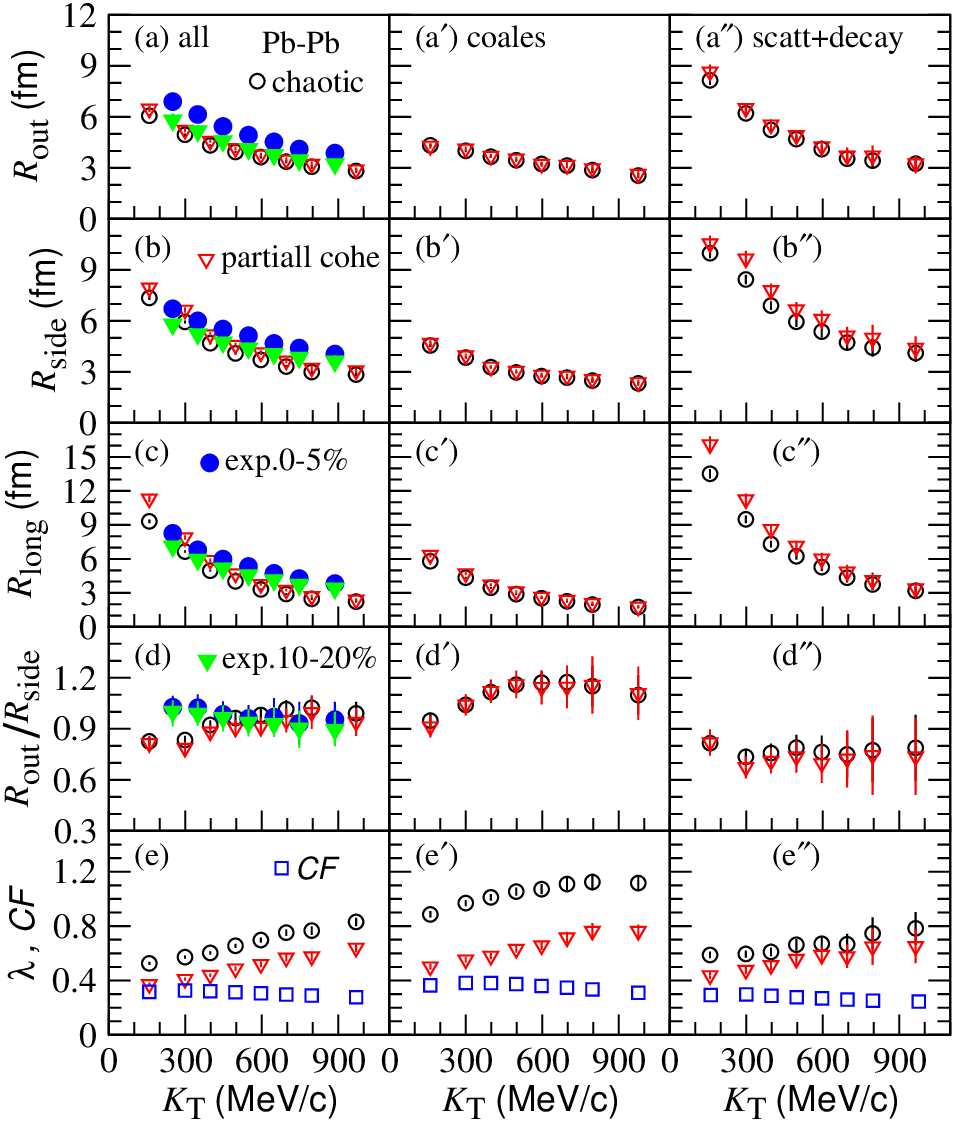}
\caption{(Color online) Two-pion HBT results ($R_{\rm out}$, $R_{\rm side}$, $R_{\rm
long}$, $R_{\rm out}/R_{\rm side}$, $\lambda$) for the chaotic and partially coherent
source and the coherent fraction ($C\!F$) of partially coherent source, with respect
to transverse momentum of pion pair $K_{\rm T}$, in Pb-Pb collisions at
$\sqrt{s_{NN}}=2.76$~TeV with impact parameter $0<b<3$~fm in the AMPT model.
Experimental data \cite{ALICE_PRC93_2016a} in Pb-Pb collisions at
$\sqrt{s_{NN}}=2.76$~TeV with centralities 0-5\% and 10-20\% are presented for
comparison.}
\label{zf-HBTfit-pb}
\end{figure}

Figure~\ref{zf-HBTfit-pb} shows the results of the fitted HBT radii $R_{\rm out}$,
$R_{\rm side}$, and $R_{\rm long}$; ratio $R_{\rm out}/R_{\rm side}$; and
chaoticity parameter $\lambda$ for chaotic and partially coherent sources in
AMPT Pb-Pb collisions at $\sqrt{s_{NN}}=2.76$~GeV, with impact parameter $0<b<3$~fm,
in the transverse momentum intervals of pion pair $K_{\rm T}<250$~MeV$\!/c$, $250\leq
K_{\rm T}<350$~MeV$\!/c$, $350\leq K_{\rm T}<450$~MeV$\!/c$, $450\leq K_{\rm T}<550$~MeV
$\!/c$, $550\leq K_{\rm T}<650$~MeV$\!/c$, $650\leq K_{\rm T}<750$~MeV$\!/c$, $750\leq
K_{\rm T}<850$~MeV$\!/c$, and $850\leq K_{\rm T}<1200$~MeV$ \!/c$.
Figure~\ref{zf-HBTfit-pb} also shows the coherent fraction $C\!F$ of partially coherent
source in the AMPT model, and the experimental data of HBT radii in Pb-Pb collisions
at $\sqrt{s_{NN}}=2.76$~GeV with centralities 0-5\% and 10-20\% \cite{ALICE_PRC93_2016a}
for comparison. The numerical results of $C\!F$ are presented in Table II with the ratios
$n_c/n_t$. The behaviours of the ratio $n_c/n_t$ and coherent fraction $C\!F$ in the
cases for all the pions, the pions generated by quark coalescence, and the pions
generated by particle scattering and decay are similar to those in Au-Au collisions.

\begin{table*}[htb]
\begin{center}
\caption{Ratio of coherent pion-pair number $n_c/n_t$, and coherent fraction $C\!F$, in
Pb-Pb collisions at at $\sqrt{s_{NN}}=2.76$~TeV, in the AMPT model. The superscripts (1),
(2), and (3) are for all of the generated pions, the pions generated by quark coalescence,
and the pions generated by particle scattering and decay, respectively.}
\begin{tabular}{c|cc|cc|cc}
\hline\hline
Pb-Pb@2.76\,TeV&~$n_c^{(1)}/n_t^{(1)}$~&~~$C\!F^{(1)}$~~&~~$n_c^{(2)}/n_t^{(2)}$~&
~~$C\!F^{(2)}$~~&~~$n_c^{(3)}/n_t^{(3)}$~&~~$C\!F^{(3)}$~~\\
\hline
$K_T\!<\!250\,\text{MeV\!/\!c}$&0.100&0.316&0.133&0.364&0.086&0.293\\
~$250\!\!\leq\!K_T\!<\!\!350\,\text{MeV\!/\!c}$&0.106&0.326&0.146&0.382&0.089&0.298\\
~$350\!\!\leq\!K_T\!<\!\!450\,\text{MeV\!/\!c}$&0.103&0.321&0.145&0.381&0.082&0.287\\
~$450\!\!\leq\!K_T\!<\!\!550\,\text{MeV\!/\!c}$&0.099&0.314&0.140&0.374&0.077&0.277\\
~$550\!\!\leq\!K_T\!<\!\!650\,\text{MeV\!/\!c}$&0.093&0.305&0.130&0.360&0.072&0.268\\
~$650\!\!\leq\!K_T\!<\!\!750\,\text{MeV\!/\!c}$&0.088&0.297&0.121&0.347&0.068&0.260\\
~$750\!\!\leq\!K_T\!<\!\!850\,\text{MeV\!/\!c}$&0.083&0.289&0.111&0.334&0.064&0.252\\
~$850\!\!\leq\!K_T\!<\!\!1200\,\text{MeV\!/\!c}$&0.077&0.277&0.096&0.310&0.060&0.245\\
\hline\hline
\end{tabular}
\label{Tab-coh-Pb}
\end{center}
\end{table*}

In comprehensively comparing the model HBT radii $R_{\rm out}$, $R_{\rm side}$, and
$R_{\rm long}$ with the experimental data, and the $C\!F$ results with the values
extracted from experimental measurements of multi-pion HBT correlations
\cite{ALICE_PRC89_2014,ALICE_PRC93_2016}, we took the longitudinal and transverse
coherent-length parameters $a_{\rm Z}$ and $a_{\rm T}$ to be 0.8 and 2.5, respectively,
for the partially coherent source.
One can see from Figs.~\ref{zf-HBTfit-pb}(a)--\ref{zf-HBTfit-pb}(e) that the model HBT
radii of the partially coherent source are closer to the experimental data than those
of the chaotic source. Also, the results of coherent fraction of the partially coherent
source are consistent with the values 0.23$\pm$0.08 extracted from the experimental
measurement of three-pion HBT correlation functions \cite{ALICE_PRC89_2014} and
0.32$\pm$0.03(stat)$\pm$0.09(syst) extracted from the experimental measurement of
four-pion HBT correlation functions \cite{ALICE_PRC93_2016}.
The middle and right panels in Fig.~\ref{zf-HBTfit-pb} show the results for the chaotic
and partially coherent sources with the pions generated by quark coalescence, and particle
scattering and decay, respectively. One can see that the differences between the $\lambda$
values for chaotic and partially coherent sources are great in the case of quark coalescence.
However, the differences of $\lambda$ values between the chaotic and partially coherent
sources are slight in the case of particle scattering and decay.

\section{Summary and Discussion}
We performed two-pion interferometry for pion-emitting sources in relativistic
heavy-ion collisions in the AMPT model. A partially coherent source was constructed by
introducing momentum dependent longitudinal and transverse coherent emission lengths
$L_{\rm CZ}$ and $L_{\rm CT}$, based on pion de Broglie wavelength, in the calculations
of two-pion HBT correlation functions. The two pions generated with longitudinal distance smaller than $L_{\rm CZ}$ and with transverse distance smaller than $L_{\rm CT}$ were
assumed to be emitted coherently and without HBT correlation. We compared the HBT results
in the AMPT model with experimental data in Au-Au collisions at $\sqrt{s_{NN}}=200$~GeV,
and in Pb-Pb collisions at $\sqrt{s_{NN}}=2.76$~TeV.
We found that the HBT chaoticity parameter $\lambda$ decreases with increasing $L_{\rm
CZ}$ and $L_{\rm CT}$. The results of HBT radii $R_{\rm out}$, $R_{\rm side}$, $R_{\rm long}$, and chaoticity parameter $\lambda$ of partially coherent sources in the AMPT
model are closer to the experimental data in Au-Au collisions at $\sqrt{s_{NN}}=200$~GeV
at the RHIC, compared to those of chaotic sources.
Also, the results of $R_{\rm out}$, $R_{\rm side}$, and $R_{\rm long}$ of the partially
coherent sources in the AMPT model are closer to the experimental data in Pb-Pb collisions
at $\sqrt{s_{NN}}=2.76$~TeV at the LHC, compared to those of the chaotic sources.
The results of coherent fraction of partially coherent source in Pb-Pb collisions in
the AMPT model are consistent with the values extracted by experimental measurements of
multi-pion HBT correlations.
We investigated the two-pion HBT results for chaotic and partially coherent
sources constructed with the pions generated by quark coalescence and by particle
scattering and decay, and found that the chaoticity parameter values of partially coherent
sources are significantly less than those of chaotic sources for quark coalescence pions.

With the AMPT model, one can trace back a freeze-out particle to its generation
coordinates, momentum, and parent. It is useful for studying the underlying physics
of experimental observables. The work presented in this paper provides a possible
relationship between pion generation and the HBT observable, the two-pion correlation
function. However, it does not affect the explanations that the AMPT model provides
to other observables, such as single-particle spectra and collective flows.
In this study, the coherent-length parameters $a_Z$ and $a_T$ are determined by
comparing the model HBT results with experimental data. They are model dependent.
Physically, the parameters are related to the thermodynamical environment of particle
production. Investigating the dependence of the coherent parameters on source
thermodynamical properties and obtaining more general coherent parameters will be of
interest.
Pions, as the lightest hadron, may involve quantum effects in their production and
propagation in the sources. It is possible that some of these effects remain and
affect two- and multi-pion HBT correlations, the observables from quantum statistics.
In relativistic heavy-ion collisions, the suppressions of two- and multi-pion HBT
correlation functions at small relative momenta may indicate the pion-emitting sources
are partially coherent. More detailed studies of the suppressions of pion HBT correlations
in relativistic heavy-ion collisions will be of interest.

\begin{acknowledgements}
We thank Zi-Wei Lin for useful discussions and suggestions.
This research was supported by the National Natural Science Foundation of
China under Grant No. 12175031.
\end{acknowledgements}

\end{document}